\documentclass[aps,prl,twocolumn,groupedaddress]{revtex4-1}
\usepackage{bm}
\usepackage{graphicx}
\DeclareGraphicsExtensions{.pdf,.png,.eps}
\usepackage{epstopdf}

\newcommand{\te}[1]{\mbox{\boldmath $ #1 $}}

\def\bU{\te{U}}

\def\bF{\te{F}}

\def\b1{\te{1}}

\begin{document}


\title{Tracer diffusion in active suspensions}


\author{Eric W. Burkholder, John F. Brady}
\email[]{jfbrady@caltech.edu}
\affiliation{Division of Chemistry and Chemical Engineering, California Institute of Technology, Pasadena, California 91125, USA}


\date{\today}

\begin{abstract}
We study the diffusion of a Brownian probe particle of size $R$ in a dilute dispersion of active Brownian particles (ABPs) of size $a$, characteristic swim speed $U_0$, reorientation time $\tau_R$, and mechanical energy $k_s T_s = \zeta_a U_0^2 \tau_R /6$, where $\zeta_a$ is the Stokes drag coefficient of a swimmer. The probe has a thermal diffusivity $D_P = k_B T/\zeta_P$, where $k_B T$ is the thermal energy of the solvent and $\zeta_P$ is the Stokes drag coefficient for the probe. When the swimmers are inactive, collisions between the probe and the swimmers sterically hinder the probe's diffusive motion. In competition with this steric hindrance is an enhancement driven by the activity  of the swimmers.  The strength of swimming relative to thermal diffusion is set by $Pe_s = U_0 a /D_P$. The active contribution to the diffusivity scales as $Pe_s^2$ for weak swimming and $Pe_s$ for strong swimming, but the transition between these two regimes is nonmonotonic. When fluctuations in the probe motion decay on the time scale $\tau_R$, the active diffusivity scales as $k_s T_s /\zeta_P$: the probe moves as if it were immersed in a solvent with energy $k_s T_s$ rather than $k_B T$.
\end{abstract}


\maketitle

Diffusive and rheological properties of active suspensions are important for understanding many biological systems and processes, such as  transport within cells.  Active Brownian particles (ABPs), which move with a self-propulsive velocity $\bU_0$  and randomly reorient with a characteristic time scale $\tau_R$, provide a minimal model for active suspensions; even the precise mechanism of their autonomous motion need not be specified.  The motion of these active particles, or ``swimmers,''  affects not only material properties (e.g. viscosity), but also the motion of passive constituents, such as nutrients or signaling proteins that may be important for cell survival. 

In a passive suspension where  particles lack the ability to self-propel, it is well known that  ``collisions'' between a probe and the bath particles sterically hinder the long-time diffusive motion of a probe; the effective long-time diffusivity is less than the isolated Stokes-Einstein-Sutherland (SES) value  \cite{Batchelor1976, Zia2010}.  By contrast, experiments have confirmed that colloidal tracers (both Brownian and non-Brownian) in active bacterial suspensions undergo enhanced diffusive motion at long times due to bath activity. This is observed not only in liquid cultures, but also in porous media and on agar surfaces \cite{Kim2004, Wu2000,Wu2011}. As a result,  recent theoretical and experimental investigations have been motivated to understand the character of this enhanced diffusive motion and to provide models that describe this  behavior \cite{Jepson2013,Mino2011,Mino2013, Morozov2014,Kasyap2014,Lin2011,Thiffeault2010}. For example, Kasyap et al. \cite{Kasyap2014} developed a mean-field hydrodynamic theory to describe the effects of binary interactions between point tracers and ellipsoidal bacterial swimmers. This theory predicts a net enhancement of  tracer diffusivity arising from the fluid flow induced by the swimming bacteria, which was shown to be a nonmonotonic function of a P\'{e}clet number relating the strength of bacterial advection to the Brownian motion of the tracer.  Experimental studies have also observed a nonmonotonicity in P\'{e}clet number when varying the size of the tracer particle \cite{Patteson2016}. Other theory and experiments propose that the  enhancement to the diffusivity is linear in the ``active flux" due to the swimmers' autonomous motion \cite{Jepson2013,Mino2011,Mino2013, Morozov2014}. 

Here we show that these same qualitative features are recovered without considering hydrodynamic interactions (HI)---the enhanced diffusivity of passive particles may be understood as a result of the activity of the bath particles and excluded volume interactions alone. This does not mean the HI are not important, only that their effect is quantitative, not qualitative.  We use  a Smoluchowski-level analysis to model the active suspension and compute the long-time diffusivity of a passive probe using generalized Taylor dispersion theory and expansions in orientational tensor harmonics \cite{Yan2015,Zia2010,Saintillan2015}. The derivation and complete expressions for the active diffusivity of the probe are given in the supplemental material \cite{SuppInfo2017}; here we focus on limiting behaviors. Additionally, we show that these excluded volume interactions have important implications for experimental measurements of activity-enhanced diffusion:  steric hindrance to passive diffusion is in competition with active enhancement and  both effects must be considered when designing and analyzing experiments. 

Consider a passive Brownian particle of size $R$ moving through a bath comprised of a Newtonian solvent of viscosity $\eta$, and a dispersion of ABPs of size $a$, swim speed $U_0$, and reorientation time $\tau_R$. In the absence of the probe, the swimmers undergo both a thermal and an active random-walk, where the thermal  walk is characterized by the SES diffusivity $D_a$, and the random walk due to their self-propulsion is characterized by a swim diffusivity $D^{swim} = U_0 ^2 \tau_R /6$.  We define the mechanical activity of the bath as the Stokes drag times the swim diffusivity: $k_s T_s = \zeta_a D^{swim}$, just as  $k_B T = \zeta_a D_a$ \cite{Takatori2014, Takatori2014b}.  The volume fraction of swimmers is $\phi = 4\pi a^3 n^\infty/3$, where $n^\infty$ is the uniform number density of swimmers far from the probe.
The probe has a thermal diffusivity $D_P = k_B T/\zeta_P$, and the probe-swimmer pair has a relative thermal diffusivity $D^{rel} = D_a + D_P$. The competition between swimming and Brownian motion is governed by the swim P\'{e}clet number: $Pe_s = U_0 R_c /D^{rel} =U_0 R/D_a = U_0 a/D_P$,  and $R_c = R + a$ is the  center-to-center separation distance of the probe and swimmer upon contact.

In the absence of activity, the (passive) bath particles hinder the probe's motion due to steric interactions \cite{Batchelor1976}. For dilute suspensions the active contribution to the diffusivity is
$\langle \bm{D^{act}} \rangle \equiv \langle \bm{D^{eff}} \rangle  -D_P \bm{I}(1-\phi_{act}),$
where $\langle \bm{D^{eff}} \rangle$ is the effective diffusivity of the probe and $\phi_{act} \equiv \phi (R_c/a)^2 /2$ measures the number of swimmers colliding with the probe (which can be much larger than the actual volume fraction $\phi$ for large probes). 
The diffusivity of a probe in a suspension of inactive swimmers is $D_P \bm{I} (1-\phi_{act})$.  When the probe and  ABP are the same size, $\phi_{act} = 2\phi$, and the steric reduction is $1 - 2\phi$, a well-known result in the absence of HI \cite{Batchelor1976}.  
Both the effective and active diffusivities are isotropic. 

We can predict $D^{act}$ with simple scaling arguments. The kinematic definition of the diffusivity is $D^{act} = N (U^\prime)^2 \tau$, where $U^\prime$ is the magnitude of the probe's velocity fluctuations due to collisions with the swimmers, $\tau$ is the time scale over which these fluctuations become decorrelated, and $N$ is the number of swimmers  colliding with the probe. Upon collision a swimmer  pushes the probe with its propulsive swim force ${\bF^{swim}} = \zeta_a {\bU_0}$, while the solvent  resists this motion via the probe's Stokes drag. Thus, the magnitude of velocity fluctuations is $U^\prime \sim \zeta_a U_0 / \zeta_P$. (When the probe is small compared to the swimmers, the velocity fluctuations scale with the swim speed, $U^\prime \sim U_0$.) On average the probe will experience $N \sim n^\infty R_c ^3$ collisions, where  $R_c^3$ is the volume occupied by a swimmer-probe pair.  Hence,
\begin{eqnarray}
D^{act} & \sim &  n^\infty R_c^3 \left(\frac{\zeta_a}{\zeta_P}\right)^2 U_0 ^2 \, \tau, \quad R\gtrsim a, \nonumber\\*
 &  & n^\infty R_c^3 \, U_0^2 \, \tau , \quad  R \ll a \, .
 \label{eq:scaling}
\end{eqnarray}
The  time scale $\tau$ differs depending on the dominant physical process governing the decorrelation and can take one of three values: (1) the diffusive time $\tau_D = R_c^2/D^{rel}$, (2)  the advective time $\tau_{adv} = R_c/U_0$ and (3) the reorientation time $\tau_R$.  

\begin{figure}
\centering
\includegraphics[width = \linewidth]{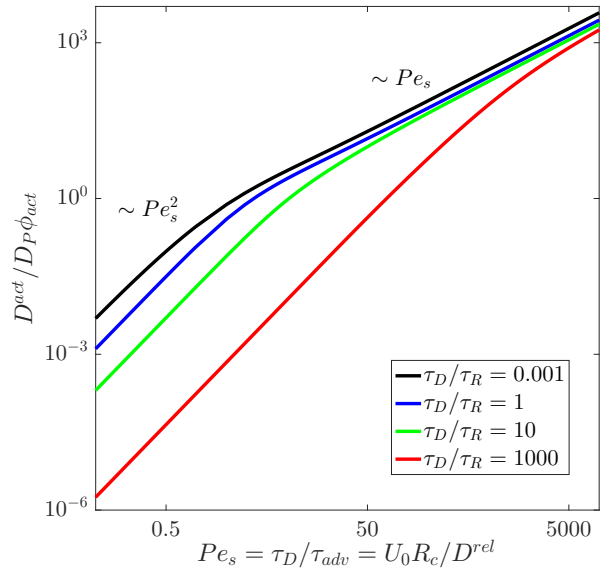}
\caption{Active diffusivity of the probe as a function of the ratio of the pair-diffusion time to the advection time $Pe_s = \tau_D/\tau_{adv} = U_0 R_c / D^{rel}$, where $U_0$ is the swim speed, $R_c$ is the center-to-center separation distance of the probe and swimmer upon contact, and $D^{rel}$ is the relative thermal diffusivity of the probe-swimmer pair. The ratio $\tau_D/\tau_R$ indicates the strength of Brownian motion relative to the reorienations of the swimmers. The active diffusivity is non-dimensionalized by the probe's SES diffusivity $D_P$ times the active volume fraction $\phi_{act} = (4 \pi/3)n^\infty R_c ^2 a/2$, where $a$ is the swimmer size and $n^\infty$ is the number density of swimmers. }
\label{fig:peclet}
\end{figure}

(1) When the decorrelation time $\tau = \tau_D \equiv R_c^2/D^{rel}$, the probe's fluctuations are induced by the swimming bath particles, but the fluctuations are sufficiently weak ($Pe_s \ll 1$) that they decay on the time scale of Brownian diffusion. The scaling argument predicts $D^{act} \sim D_P Pe_s ^2 \phi_{act}$, and the detailed calculations give
\begin{equation}
D^{act} = \frac{29}{54}D_P Pe_s^2 \phi_{act},
\label{eqn:pe2}
\end{equation}
as one would expect for Taylor dispersion:  the linear response diffusivity scales as $Pe_s ^2$ (or $U_0^2$).   Kasyap et al.~\cite{Kasyap2014} found that the hydrodynamically-driven diffusivity of a point tracer scales as $Pe_s^{3/2} \sqrt{U_0 \tau_R /a}$ when swimming is weak, which is also quadratic in $U_0$.  We predict that $D^{act} \sim Pe_s^2$ for all $a/R$, but curiously we find no explicit  dependence on $\tau_R$, although such a dependence is evident in Fig.~\ref{fig:peclet}; we address this in (3) below.

(2) When swimming is strong compared to Brownian motion, the appropriate time scale is $\tau = \tau_{adv} = R_c/U_0$. The swimmers are bombarding the probe so rapidly that the resulting fluctuations become decorrelated on the time it takes for a swimmers to traverse the distance $R_c$. The scaling analysis (\ref{eq:scaling}) predicts  $D^{act} \sim D_P Pe_s \phi_{act} \sim U_0 a \phi_{act}$, and the detailed Smoluchowski approach gives:
\begin{equation}
D^{act} = \frac{1}{3 \sqrt{3}} U_0 a \left(\frac{2+\sqrt{2 \tau_D/\tau_R}}{1+\sqrt{2 \tau_D/\tau_R}}\right)\phi_{act}.
\end{equation}
The probe's diffusivity is now linear in the swim speed $U_0$ (or linear in $Pe_s$), as expected from Taylor dispersion theory.  Kasyap et al.~\cite{Kasyap2014} find that $D^{act} \sim n^\infty a^3 U_0 a$ (because the tracers have no size in their analysis the only geometric length scale is the swimmer size $a$), but their result is independent of $\tau_R$.  The transition from diffusive to advective behavior is shown in Fig.~\ref{fig:peclet}.

In this limit the run length of a swimmer, $\ell \equiv U_0 \tau_R$, is large compared to the pair size $R_c$, and a swimmer  collides with the probe before it is able to traverse its full run length. The swimmer pushes the probe with force $\zeta_a U_0$, but is only able to move it a distance of $O(a)$ on average. One might think that the swimmer should be able to push the probe the contact length $R_c$, but the no-flux boundary condition allows the swimmer to slide along the probe's surface, and thus the average distance of a push is only $O(a)$. 
Just as in the diffusion-controlled regime, the result is insensitive to the swimmer-probe size ratio $a/R$. It manifests only in $\phi_{act}$, which simply becomes $\phi$ for point tracers. Finally, we note that the ratio of the other two time scales $\tau_D/\tau_R$ has no bearing on the scaling of the diffusivity in this limit---it can only change the result by a factor of two. 

However, $\tau_D/\tau_R$ significantly affects the behavior in the diffusion-dominated regime and the location of the transition from the diffusive to advective behavior.  When $\tau_D/\tau_R \ll 1$, reorientations are slow and the transition occurs for $Pe_s \sim O(1)$ as one would expect. However, as reorientations become faster ($\tau_D/\tau_R$ increases), the transition occurs at much higher values of $Pe_s$ (see Fig.~\ref{fig:peclet}). In the athermal  limit of no translational diffusion ($\tau_D \rightarrow \infty$), the transition to strong swimming is governed  by the reorienation P\'{e}clet number $Pe_R \equiv \tau_{adv} /\tau_R = R_c/\ell \sim O(1)$ rather than the swim P\'{e}clet number $Pe_s$.

(3) When Brownian motion is weak compared to the swimmers' reorientations, the decorrelation time is set by the reorientation time:  $\tau = \tau_R$. The scaling arguments predict $D^{act} \sim (k_s T_s / \zeta_P) \phi_{act}$, or $D^{act} \sim D^{swim}\phi$ for small probes.
The result of the Smoluchowski analysis is in agreement:
\begin{equation}
D^{act} = \left(\frac{k_s T_s}{\zeta_P}\right)\frac{R}{R_c}\phi_{act}\, .
\end{equation}
Note that  there is no dependence on  $k_B T$. 

Suppose that the swimmers and probe are  large enough so that Brownian motion is not important, but the swimmers' reorientation time is relatively fast. The probe receives many small active kicks of size $k_s T_s$ from the  swimmers, which are dissipated by the Stokes drag $\zeta_P$. Thus, the diffusivity looks like what one would expect from a stochastic ``Brownian" process, where the energy is $k_s T_s$ rather than $k_B T$.  In the limit when the probe is very small, $(k_s T_s /\zeta_P)(R/R_c) \rightarrow U_0 ^2 \tau_R /6$, $\phi_{act} \rightarrow \phi$, and the active diffusivity is simply the swim diffusivity times the volume fraction of swimmers: $D^{act} = D^{swim} \phi$. As a swimmer hops in one direction and equal volume for solvent is displaced in the opposite direction. 

Because the probe receives many small kicks from the swimmers, its motion is governed by a Langevin equation 
$0 = -\zeta_P \bm{U} + \bm{F}^{swim},$
where $\bm U$ is the probe velocity and the swimmers exert a fluctuating  force with  zero mean $\langle \bm{F}^{swim} \rangle = \bm 0$ and autocorrelation $\langle \bm{F}^{swim}(t)\bm{F}^{swim}(t')\rangle = 2 k_s T_s \zeta_P \bm{I} \delta(t-t')$ for times long compared to $\tau_R$. The mean-squared displacement follows as $\langle (\Delta \bm{x}(t))^2\rangle = 2 (k_s T_s/\zeta_P) t \bm{I}$ for the diffusivity of a particle immersed in such an active medium. 

\begin{figure}
\centering
\includegraphics[width =\linewidth]{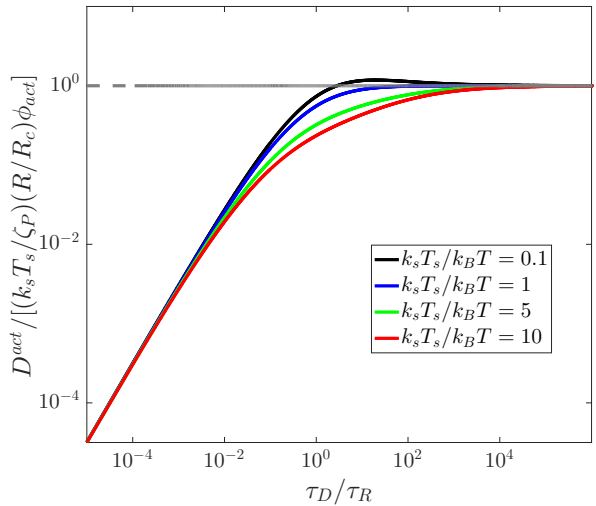}
\caption{Active diffusivity of the probe non-dimensionalized by $(k_s T_s /\zeta_P)(R/R_c) \phi_{act}$ as a function of the ratio of the diffusion time to the swimmer reorientation time $\tau_D /\tau_R = R_c ^2 / \tau_R D^{rel}$ for various values of the mechanical to thermal energy, $k_s T_s/k_BT$, where $ k_s T_s= \zeta_a U_0^2 \tau_R/6$. 
}
\label{fig:continuum}
\end{figure}

In this ``continuum limit" the probe acts as a thermometer that measures the swimmers' activity  $k_s T_s$. When $\ell/R_c \rightarrow 0$, active suspensions have a well-defined `temperature' through their activity $k_s T_s$ \cite {Takatori2015} because the motion looks like a stochastic Brownian process. When $\ell/R_c \gg 1$, as is the case in the strong swimming regime, the definition of temperature breaks down because the swimmers no longer move the probe a distance $\ell$, they only push it a distance $a$ between reorientations. Thus, the swimmers do not ``share" their activity fully with the probe; the appropriate shared quantity in this limit is $Pe_R$. 

\begin{figure}
\centering
\includegraphics[width =\linewidth]{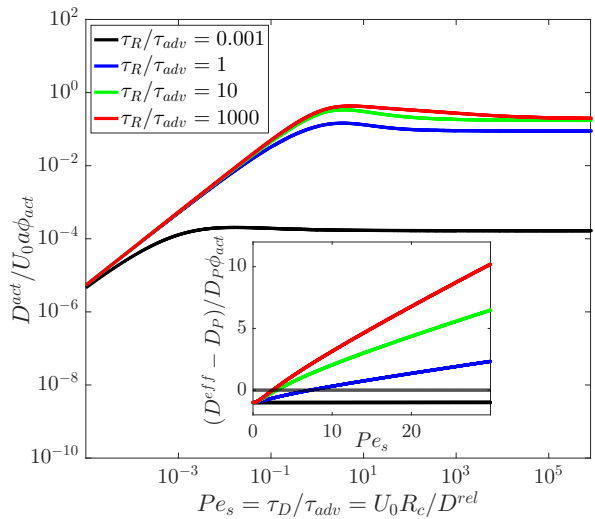}
\caption{Active diffusivity of the probe non-dimensionalized by $U_0 a$ as a function of $Pe_s = \tau_D/\tau_{adv} = U_0 R_c / D^{rel}$. 
The ratio $\tau_R/\tau_{adv} = U_0\tau_R/R_c = \ell/R_c$ reflects the speed of reorientation relative to advection. The inset shows the total $O(\phi_{act})$ change in the probe's diffusivity, non-dimensionalized by $D_P \phi_{act}$, where $D_P$ is the bare diffusivity of the probe. }
\label{fig:nonmonotonic}
\end{figure}

Figure \ref{fig:continuum} shows $D^{act}$ as a function of $\tau_D/\tau_R$ for various values of $D^{swim}/D^{rel} = (\tau_D/\tau_R)/\tau_{adv}^2 \sim k_s T_s/k_B T$. For $\tau_D/\tau_R \rightarrow \infty$ we recover the continuum-like scaling for any value of $k_s T_s /k_BT$. Though intuition might say that the diffusivity should be dominated by thermal kicks when $k_s T_s \ll k_B T$, it is important to remember that it is the solvent, not the bath particles, that give the probe thermal kicks. The swimmers can only give kicks of size $k_s T_s$. The finite size of the swimmers replaces a volume of solvent, thus reducing the number of thermal kicks the probe receives. The $O(\phi_{act})$ change in the probe diffusivity is actually {\em negative} when $k_s T_s  < k_B T$ (see the inset of Fig.~\ref{fig:nonmonotonic}): steric hinderance exceeds active enhancement.

An interesting feature predicted by the detailed theory is a nonmontonic dependence of $D^{act}$ on both $\tau_D/\tau_R$ and $Pe_s$, as seen in Figs~\ref{fig:continuum} and \ref{fig:nonmonotonic}, respectively.  As $Pe_s$ increases, thermal diffusion slows and swimming becomes more important, so we transition from a diffusive to advective behavior. This transition does not occur monotonically with $Pe_s$ because $Pe_R = \tau_{adv}/\tau_R$ also influences the dynamics. Imagine a scenario where $\tau_D$ and $\tau_R$ are fixed and $R\gg a$, but we adjust the swimmers' speed (perhaps by altering the amount of available fuel). When the swimmers move slowly, Brownian motion dominates: $\overline{D^{act}} \equiv D^{act}/(U_0 a \phi_{act}) \sim Pe_s$. When the swim speed is large, advection dominates and $\overline{D^{act}}$ is constant.  When $\tau_D \sim \tau_{adv}$, neither wins out and the reorientations are allowed to influence the dynamics. Finite Brownian motion keeps the swimmers close to the probe after a collision, and slow reorientation allows the swimmer to collide with the probe again rather than run off, thus the diffusivity is slightly higher than the advective scaling. When reorientations are too fast, this peak dissapears. This is corroborated by Fig.~\ref{fig:continuum}, which reveals that $D^{act}$ is only nonmonotonic when $k_s T_s < k_B T$. The nonmontonicity still occurs when $\tau_D \sim \tau_R$, but Brownian motion is only strong enough to compete with activity if the thermal energy of the solvent exceeds the activity of the bath. 

Kasyap et al.~\cite{Kasyap2014} find the same phenomenon in their treatment. When the diffusion is hydrodynamic in origin and advection dominates, the tracer follows a straight trajectory along fluid streamlines. Weak Brownian motion allows the tracer to sample more trajectories, and the odd symmetry of the bacterium's dipolar flow field results in an increased correlation in probe motion. When Brownian motion is strong, the probe's motion decorrelates and the diffusivity decreases. Thus the diffusivity decreases nonmonotonically with increasing Brownian motion (i.e. as one moves from right to left in Fig.~\ref{fig:nonmonotonic}). Patteson et al.~\cite{Patteson2016} see something similar in experiments by varying the probe size, which is equivalent to varying $Pe_s$ when all other parameters are fixed. They scale $D^{act}$ by $n^\infty L^3 U_0 L$, where $L$ is the total bacterium length. They find that this scaled diffusivity first increases with probe size as approximately $R^2$ and then decreases to a plateau. Our scaling analysis predicts that $\overline{D^{act}}$ is linear in probe size when diffusion dominates, and indepedent of probe size when advection dominates. In between, when the appropriate time scale is $\tau_R$, $\overline{D^{act}}$ scales as $1/R$, thus capturing the nonmonotonicity. The peak in $\overline{D^{act}}$ is predicted around $Pe_s \sim 5$ in our study and in \cite{Kasyap2014}, but is found experimentally around $Pe_s \sim O(10^3)$; the source of such a large discrepancy is not known. Lastly, we note that the inset of Fig.~\ref{fig:nonmonotonic} shows that this nonmonotonicity is obscured by the steric hindrance, reinforcing the importance of considering excluded-volume interactions in active suspensions.

Another common model, used by Mi\~{n}o et al.~\cite{Mino2011} to describe enhanced diffusion of tracers in bacterial suspensions, says that the active enhancement is proportional to the advective flux of the active particles: $D^{eff} = D_P + \beta J_a$, where $J_a = n^\infty U_0$ in our notation, similar to what we find for strong swimming. 
Lin et al.~\cite{Lin2011} predict that $\beta$ scales as the body size to the fourth power for squirmers, but subsequent theoretical derivations indicate that $\beta^{1/4}$ also depends on the swimmer's hydrodynamic dipole moment, particle size, system geometry, swimming efficiency, etc. As in \cite{Kasyap2014}, these studies do not take the swimmers to be thermally active. Additionally, they argue that the size of the tracer particle does not affect $\beta$ \cite{Mino2011}, and thus excluded-volume effects are generally neglected.  This is valid when the tracer particles are always far enough away from the bacteria that the size effects in the Fax\'{e}n expression for their velocity are negligible, which is consistent with theoretical models that assume the bacteria to be simple hydrodynamic dipoles (which is only true in the far field \cite{Drescher2010,Drescher2011}). 

For this $\beta$ model, our Smoluchowski theory predicts
$\beta = (2 \pi / 9\sqrt{3})R_c^2 a^2[(2+\sqrt{2 \tau_D/\tau_R})/(1+\sqrt{2 \tau_D/\tau_R})]$.
The ability of the swimmer to randomly reorient is not required for this enhancement to the diffusivity, as argued in \cite{Mino2013}. In contrast to some of these experimental studies, our result depends on the size of the tracer particle. In the system of Jepson et al. \cite{Jepson2013} the tracers are non-motile \textit{E. Coli} in a suspension of motile \textit{E. Coli} with equivalent spherical dimension $a = 1.4 \mu m$. From their experimental parameters, we predict $\beta = 3.22 a^4 - 6.45 a^4$. To match the experimentally found value of $\beta = 7.1 \mu m^4$, our theory predicts that the \textit{E. Coli} would have an equivalent spherical dimension of $a = 1.02 - 1.22 \mu m$. 

As previously proposed, this advective flux model ignores the steric hinderance of the passive suspension, which should accounted for by 
\begin{equation}
D^{eff} = D_P(1-\phi_{act}) + \beta J_a\, .
\end{equation}
The steric hinderance is especially important when swimming is weak (Fig.~\ref{fig:nonmonotonic}). 
Experimentally, one should measure the bare diffusivity of a tracer, and then the change in diffusivity among non-motile swimmers to recover the effective particle size $R_c$ from Batchelor's theory \cite{Batchelor1976}.  Knowing $R_c$, the average swim speed, reorienation time, and the bare particle diffusivities, one  can calculate the active diffusivity from our theory, and then compare to experimental measurements.

We presented a micromechanical model for the effective diffusivity of a passive particle embedded in a suspension of ABPs. Using a generalized Taylor dispersion approach, and employing an expansion in orientational tensor harmonics, we found an exact analytical expression for the effective diffusivity of a Brownian probe for arbitrary particle sizes, swimmer activity, and time scales (\cite{SuppInfo2017}). Our theory agrees qualitatively with previous experimental and theoretical investigations of enhanced diffusion in active suspensions, and is able to explore regimes of parameter space not typically considered in most experiments. It highlights several key features of diffusion in active suspensions: (1) the diffusion of a tracer is nonomontonic in a P\'{e}clet number comparing swimming to thermal diffusion, (2) steric hindrance of tracer motion is in competition with the enhancement due to bath activity, and (3), when fluctuations of the tracer's motion decorrelate on the same time scale as swimmers' reorientations, the bath mimics a homogenous solvent with energy $k_s T_s$.

\begin{acknowledgments}
This work is funded by NSF grant no. CBET 1437570. We thank S.C. Takatori for helpful discussions. 
\end{acknowledgments}

\bibliography{library}
\end{document}